\documentclass[smus]{snow2e_d}
\usepackage{epsfig}

\begin{document}
\hyphenation{counter-terms}

\begin{titlepage}

\headsep 2cm
\baselineskip 0.5cm

\evensidemargin 0.7cm

\begin{flushright}
{\large FERMILAB-CONF-96/424-T \\
UB-HET-96-04}
\end{flushright}

\vspace{1.5cm}

\begin{center}
{\Large \bf Electroweak Radiative Corrections to $W$ Boson Production
at the Tevatron}\footnote{\large To appear in the {\it Proceedings of
the 1996 DPF/DPB Summer Study on New Directions for High-Energy Physics}
(Snowmass 96)}

\vspace{2cm}

{\Large U.~Baur} \\
\vspace{0.2cm}
{\large \it Physics Department, SUNY at Buffalo, \\
Buffalo, NY 14260, USA} \\
\vspace{1cm} 
{\Large S.~Keller and D.~Wackeroth} \\
\vspace{0.2cm}
{\large \it Fermilab, P.O.~Box 500, \\
Batavia, IL 60510, USA}

\vspace{6cm} 

{\Large \bf Abstract}\\
\vspace{0.5cm}
{\large We present some results of a new calculation of the ${\cal O}(\alpha)$
electroweak radiative corrections to \\ 
$W$ boson production at hadron colliders with special emphasis on the
transverse mass distribution.}
\end{center}

\end{titlepage}

\title{Electroweak Radiative Corrections to $W$ Boson Production
at the Tevatron}

\author{U.~Baur \\ {\it Physics Department, SUNY at Buffalo, Buffalo, 
NY 14260, USA} \\[1.mm]
S.~Keller, D.~Wackeroth \\ {\it Fermilab, P.O. Box 500,
Batavia, IL 60510, USA}}
 
\maketitle

\thispagestyle{plain}\pagestyle{plain}

\begin{abstract} 
We present some results of a new calculation of the ${\cal O}(\alpha)$
electroweak radiative corrections to 
$W$ boson production at hadron colliders with special emphasis on the
transverse mass distribution.
\end{abstract}

\section{Introduction}
Despite the remarkable success of the Minimal Standard Model (MSM) 
in describing elementary particle interactions  
at presently accessible energies, there is little direct experimental 
information on the mechanism which
generates mass for the $W$ and $Z$ bosons.
In the MSM, spontaneous electroweak symmetry breaking is responsible for mass
generation. The existence of the Higgs boson in the MSM is 
a direct consequence of this mechanism. 

Complementary to the direct Higgs boson search at colliders ($M_H > 
58.4$~GeV~\cite{higgs} from LEP1, where $M_H$ is the Higgs boson mass), 
indirect information on $M_H$ can be extracted by confronting 
theoretical predictions for radiative corrections to electroweak observables
with high precision measurements.
Assuming that the MSM is valid, 
a global fit to the currently available data from LEP and
SLC with $\alpha_s$, the top quark mass ($m_t$) and $M_H$ 
as free parameters yields $M_H = 146^{+112}_{-68}$ GeV~\cite{wolf}.
Similar results have been obtained in Ref.~\cite{ewwg}.
The indirect constraints on $M_H$ are expected to improve considerably 
in the future with more precise measurements of the top quark mass
and the $W$ boson mass ($M_W$). Presently, their world averages are 
$m_t=175\pm 6$~GeV~\cite{top} and $M_W=80.356\pm 0.125$~GeV~\cite{wmass}.
The precise measurement of $M_W$ is therefore
one of the priorities of future collider experiments. 
LEP2 and RunII at Fermilab ($\int\!{\cal L}dt=2$~fb$^{-1}$) are aiming 
for an uncertainty on 
$M_W$ of about 40 MeV~\cite{lep} and 35 MeV (per experiment)~\cite{teva},
respectively. 
Further upgrades of the Tevatron accelerator complex (TeV33) could yield 
an overall integrated luminosity of ${\cal O}(30~\mbox{fb}^{-1})$,
and a precision of $M_W$ of about 15 MeV~\cite{tevah}.
Finally, it may be possible to measure $M_W$ at the LHC with an accuracy
better than 15~MeV, if an integrated luminosity of 10~fb$^{-1}$ is
accumulated with the accelerator operating at a reduced 
luminosity of about $10^{33}\,{\rm cm}^{-2}\,{\rm s}^{-1}$ (see 
Ref.~\cite{KW96}).

Obviously, to measure $M_W$ with high precision, 
it is crucial to fully control higher order electroweak (EW) and 
QCD radiative corrections.   

In this contribution we present some results of a new calculation of the 
EW ${\cal O}(\alpha)$ corrections to
$W$ boson production in hadronic collisions. In particular, we study the
effect of these corrections on the $W$ transverse mass
($M_T(\ell\nu)$) distribution from which $M_W$ is extracted
at the Tevatron. In a previous calculation~\cite{berends},
only the final state photonic corrections had been included, 
using an approximation in which the sum of the virtual and soft
part is indirectly estimated
from the inclusive ${\cal O}(\alpha^2)$ $W\to\ell\nu(\gamma)$ width and 
the final state hard bremsstrahlung contribution.
Our calculation includes both initial and 
(complete) final state corrections, as well as their interference.
As a result of our calculation, the current 
systematic uncertainty of $\Delta M_W=20$ MeV~\cite{syst,D0Wmass}
originating from EW radiative corrections will be reduced. We shall
only discuss the $W\to e\nu_e$ decay channel here. 
More details and a discussion of the $W\to\mu\nu_\mu$ decay channel will be 
presented elsewhere~\cite{paper}.

\section{The ${\cal O}(\alpha)$ contribution to resonant $W$ production}
When calculating the EW radiative corrections to resonant $W$ boson production
the problem arises how to treat an unstable charged gauge boson
consistently in the framework of perturbation theory.
This problem has been studied in Ref.~\cite{dw} with 
particular emphasis on finding a gauge invariant decomposition of the 
EW ${\cal O}(\alpha)$ corrections into a QED-like and a modified weak part. 
Unlike the $Z$ boson case, the Feynman diagrams which involve
a virtual photon do not represent a gauge invariant subset here. 
In Ref.~\cite{dw}, it was demonstrated that gauge invariant 
contributions can be extracted from the infrared (IR) singular
virtual photonic corrections. 
These contributions can be combined with the real photon corrections
in the soft photon region to form gauge invariant QED-like contributions
corresponding to initial state, final state and interference 
corrections. The soft photon region is defined by requiring that the
photon energy in the parton center of mass frame,
$E_{\gamma}$, is smaller than a cutoff 
$\Delta E=\delta_s\sqrt{\hat s}/2$, where $\sqrt{\hat s}$ is the 
parton center of mass energy.
In this phase space region, the soft photon approximation can be used
to calculate the cross section as long as $\delta_s$ is
sufficiently small. 
Throughout the calculation the soft singularities have been
regularized by giving the photon a fictitious mass.
In the sum of the virtual and soft photon
terms the unphysical photon mass dependence cancels, and
the QED-like contributions are IR finite. 
 
The IR finite remainder of the virtual photonic corrections and the weak  
one-loop corrections can be combined to separately gauge invariant 
modified weak contributions to the $W$ boson production and decay process.
Both the QED-like and the modified weak contributions can 
be expressed in terms of form factors, $F_{Q\!E\!D}^{a}$ and $\tilde 
F_{weak}^{a}$, which multiply the Born cross section~\cite{dw}. The 
superscript $a$ in the form factors denotes the initial state, final state or 
interference contributions.

The complete ${\cal O}(\alpha^3)$ parton level cross 
section of resonant $W$ production via the Drell-Yan mechanism
$q_i \overline{q}_{i'}\rightarrow f f'(\gamma)$ can then be written as 
follows~\cite{dw}:
\begin{eqnarray}
\label{wres}
\nonumber
\lefteqn{d \hat{\sigma}^{(0+1)} = 
d \hat \sigma^{(0)}\; [1+ 2 {\cal R}e (\tilde F_{weak}^{initial}+
\tilde F_{weak}^{final})(M_W^2)]} 
\\
& & +\sum_{a=initial,final,\atop interf.} [d\hat\sigma^{(0)}\; 
F_{Q\!E\!D}^a(\hat s,\hat t)+
d \hat \sigma_{2\rightarrow 3}^a] \; ,
\end{eqnarray}
where the Born cross section, $d \hat \sigma^{(0)}$, 
is of Breit-Wigner form
and $\hat s$ and $\hat t$ are the usual Mandelstam variables in 
the parton center of mass frame. 
The modified weak contributions have to be evaluated at $\hat 
s=M_W^2$~\cite{dw}. The IR finite contribution 
$d\hat\sigma_{2\rightarrow 3}^a$ describes 
real photon radiation with $E_{\gamma}>\Delta E$.
 
Additional singularities occur when the photon is emitted collinear
with one of the charged fermions.
These collinear singularities have been regularized by
retaining finite fermion masses. 
Thus, both $d \hat \sigma^{a}_{2\rightarrow 3}$ and $F_{Q\!E\!D}^{a}$ 
($a=initial, final$) contain large mass singular logarithms which have 
to be treated with special care.
In the case of final state photon radiation, the mass singular
logarithms cancel when inclusive observables are considered (KLN theorem).
For exclusive quantities, however, these logarithms can result in large 
corrections, and it may be necessary to perform a resummation of the soft 
and/or collinear photon emission terms.
For initial state photonic corrections, 
the mass singular logarithms survive. These terms are universal to all 
orders in perturbation theory, and can therefore be cancelled by 
universal collinear counterterms generated by `renormalizing' the 
parton distribution functions (PDF), in complete analogy to gluon 
emission in QCD.

To increase the numerical stability, it is advantageous 
to extract the collinear part from $d \hat \sigma^{a}_{2\rightarrow 3}$ 
for both the initial and final state,
and perform the cancellation of the mass singular logarithms analytically.
The collinear region is defined by requiring that the
angle between the fermion and the emitted photon is smaller than
a cutoff parameter $\delta_{\theta}$.
The reduced $2 \rightarrow 3$ contribution, 
which comprises the real photon contribution away from the
soft and collinear region, is 
evaluated numerically using standard Monte Carlo techniques.
Our method is very similar to the phase space slicing method
described in Ref.~\cite{GGBOO}.

It should be noted that the analytic cancellation of the final state 
mass singular logarithms is
only possible when realistic experimental electron identification
requirements are taken into account. This will be discussed in more
detail in Sec.~II.B.

In the remainder of this section, 
we extract the collinear behavior of $d \hat \sigma^{a}_{2\rightarrow 3}$
for both the initial and final state,  
and perform the 'renormalization' of the PDF. 

\subsection{Initial state photon radiation}
In the collinear region, for sufficiently small values of $\delta_{\theta}$, 
the leading pole approximation can be used,
and the initial state real photon contribution
can be written as a convolution of the Born cross section with a 
collinear factor (see also Ref.~\cite{steph}):
\begin{eqnarray}
\lefteqn{d \hat \sigma_{coll.}^{initial}=
\int_0^{1-\delta_s} dz \; d \hat \sigma^{(0)}}
\nonumber \\
&&\frac{\alpha}{2\pi} \Biggl\{Q_i^2 \,\left[\frac{1+z^2}{1-z} 
\log\left(\frac{\hat s}{m_i^2} \frac{\delta_{\theta}}{2}\right)
-\frac{2 z}{1-z}\right] \Biggr.
\nonumber \\
&& \Biggl. +(i\rightarrow i')\Biggr\} \; ,
\end{eqnarray}
{\it i.e.}~an incoming parton $q_i$ with momentum $p_i$, mass $m_i$ and 
charge quantum number $Q_i$ splits into a parton 
$q_i$ with momentum $z p_i$ and a photon of momentum $(1-z)p_i$. 
The value for the upper limit of the $z$ integral avoids overlapping
with the soft photon region.

The counterterms generated by the 'renormalization'
of the PDF can simply be derived from Ref.~\cite{paige}
by performing the replacement  
\[ \frac{\alpha_s}{\pi} \frac{4}{3} \rightarrow \frac{\alpha}{\pi} \;Q_i^2 
\; . \]
One then finds:
\begin{eqnarray}
\label{renormi}
\lefteqn{q_i(x,Q^2) = q_i(x) \;\left[1+\frac{\alpha}{\pi} \; Q_i^2 
\Biggl\{1-\log\delta_s -\log^2\delta_s \Biggr. \right.} 
\nonumber \\
& & \left.\Biggl. +\left(\log\delta_s
+\frac{3}{4}\right)\,\log\left(\frac{Q^2}{m_i^2}\right)
-\frac{1}{4} \lambda_{F\!C} \Biggr\} \right]
\nonumber\\ 
&& +\int_x^{1-\delta_s} \frac{d z}{z}\; q_i\left(\frac{x}{z}\right)
\; \frac{\alpha}{2 \pi} \; Q_i^2
\nonumber \\
& &  \Biggl\{\frac{1+z^2}{1-z} 
\log\left(\frac{Q^2}{m_i^2}\frac{1}{(1-z)^2}\right)
-\frac{1+z^2}{1-z}+\lambda_{F\!C} f_c \Biggr\}
\end{eqnarray}
with
\begin{equation}
f_{v+s} = 9+\frac{2 \pi^2}{3}+3\log\delta_s-2 \log^2\delta_s
\end{equation}
and
\begin{equation}
f_c = \frac{1+z^2}{1-z} \log\left(\frac{1-z}{z}\right)
-\frac{3}{2}\frac{1}{1-z}+2 z+3 \; .
\end{equation}
The $q_i(x)$ and $q_i(x,Q^2)$ are the 
unrenormalized and renormalized parton distribution functions, respectively.
The parameter $\lambda_{F\!C}$ distinguishes between the
$\overline{\mbox{MS}}$ ($\lambda_{F\!C}=0$) and the DIS scheme 
($\lambda_{F\!C}=1$). The scheme dependent contributions $f_c$ 
and $f_{v+s}$ can be derived from Ref.~\cite{reno}. $Q$ is the 
factorization scale. 

The cross section for
$p\overline p \rightarrow W (\gamma)\rightarrow l \nu (\gamma)$
is then obtained in two steps:~first, the parton level 
cross section of Eq.~(\ref{wres}) is convoluted with the unrenormalized 
PDF $q_i(x)$, and second, $q_i(x)$ is replaced by the 
renormalized PDF $q_i(x,Q^2)$ by using Eq.~(\ref{renormi}).   
The initial state QED-like contribution $F_{Q\!E\!D}^{initial}$ and 
the collinear part $d\hat \sigma_{coll.}^{initial}$,
including the effect of mass factorization, can be grouped into a single 
$2 \rightarrow 2$ contribution:
\begin{eqnarray}
\label{twoin}
\lefteqn{ d\sigma^{initial}_{2\rightarrow 2}
= \sum_{i,i'}
\int dx_1 dx_2 \; [q_i(x_1,Q^2) \; \overline{q}_{i'}(x_2,Q^2)
+(1 \leftrightarrow 2)]} \nonumber\\
& &  d\hat \sigma^{(0)} \; \frac{\alpha}{\pi}
\left\{(Q_i^2+Q_{i'}^2)\left[\left(\log\delta_s+\frac{3}{4}\right)
\log\left(\frac{\hat s}
{Q^2}\right)+\frac{\pi^2}{6} \right. \right.
\nonumber\\
& & \left. \left. -2 +\log^2\delta_s 
+ \frac{1}{4}\lambda_{F\!C} \;f_{v+s}\right]
-\log\delta_s+\frac{3}{2}+\frac{\pi^2}{24}\right\}
\nonumber\\
& &+ \sum_{i,i'} \int dx_1 dx_2 \Biggl[\int_{x_2}^{1-\delta_s} 
\frac{d z}{z}\; d\hat\sigma^{(0)} \Biggr.
\nonumber\\ 
& & \Biggl. [ Q_i^2  \, q_i(\frac{x_2}{z},Q^2) \,
\overline{q}_{i'}(x_1,Q^2)+Q_{i'}^2\,
q_i(x_1,Q^2) \,\overline{q}_{i'}(\frac{x_2}{z},Q^2)]\Biggr.
\nonumber\\
& & \Biggl. \frac{\alpha}{2\pi} \;
\Biggl\{\frac{1+z^2}{1-z}\log\left(\frac{\hat s}{Q^2}\frac{(1-z)^2}{z} 
\frac{\delta_{\theta}}{2}\right)\Biggr. \Biggr.
\nonumber\\
& & \Biggl. \Biggl. +1-z-\lambda_{F\!C} f_c \Biggr\} 
+ (1 \leftrightarrow 2) \Biggr]  \; .
\end{eqnarray}
As expected, the mass singular logarithms cancel completely.

In our calculation, we have not taken into account the QED radiative 
corrections to the Gripov-Lipatov-Altarelli-Parisi evolution of the PDF.
This introduces an uncertainty that needs to be quantified.
We will address this question in Ref.~\cite{paper}.
A study of the effect of QED on the evolution indicates that
the change in the scale dependence of the PDF is small~\cite{spies}.
To treat the QED radiative corrections in a consistent way, they
should be incorporated in the global fitting of the PDF.

\subsection{Final state photon radiation}
Similar to initial state radiation, the final state real photon corrections
in the collinear region can be described as a convolution of the Born 
cross section with a collinear factor. Using the leading pole approximation
one finds (see also Ref.~\cite{steph}):
\begin{eqnarray}
\label{fincoll}
\lefteqn{d \hat \sigma_{coll.}^{final} 
= \int_0^{1-\delta_s} d z \; d \hat \sigma^{(0)}}
\nonumber \\
& &  \frac{\alpha}{2\pi} \Biggl\{Q_f^2  \left[\frac{1+z^2}{1-z} 
\log\left(\frac{\hat s}{m_f^2} z^2 \frac{\delta_{\theta}}{2}\right)
-\frac{2 z}{1-z}\right] \Biggr.
\nonumber \\
& & \Biggl. +(f\rightarrow f')\Biggr\} \; .
\end{eqnarray}
When realistic experimental conditions are taken into account (see Sec.~III),
the electron and photon four-momentum 
vectors are recombined to an effective electron four-momentum
vector if their separation $\Delta R_{e\gamma}$ in the azimuthal 
angle -- pseudorapidity plane is smaller than a critical value $R_c$.
If the cutoff parameter $\delta_{\theta}$ is chosen to be smaller
than $R_c$ the integration over the momentum fraction $z$
in Eq.~(\ref{fincoll}) can then be performed analytically.
In this case, the mass singular logarithms
in the sum of $d\hat \sigma_{coll.}^{final}$
and the QED-like contribution $F_{Q\!E\!D}^{final}$ explicitly cancel, 
and one obtains:
\begin{eqnarray}
\label{EQ:FINAL}
\lefteqn{d\sigma^{final}_{2\rightarrow 2} = \sum_{i,i'} \int dx_1 dx_2 
\;[q_i(x_1,Q^2) \; \overline{q}_{i'}(x_2,Q^2)+(1\leftrightarrow 2)]
} \nonumber \\ 
& & d\hat \sigma^{(0)} \; \frac{\alpha}{\pi} 
\left\{-\log\delta_s+\frac{3}{2}+\frac{\pi^2}{24}+
(Q_f^2+Q_{f'}^2)\left[\frac{5}{4}-\frac{\pi^2}{6} \right. \right.
\nonumber\\
& & \left. \left.  -\left(\log\delta_s+
\frac{3}{4}\right)\log\left(\frac{\delta_{\theta}}{2}\right)
\right] \right\}\; .
\end{eqnarray}
The approximation~\cite{berends} used so far in modeling the EW
radiative corrections to $W$ boson production at the Tevatron
differs from our calculation only in the $2\rightarrow 2$ contribution.
At the par-\\ \noindent
ton level,
the difference between Eq.~(\ref{EQ:FINAL}) and the approximation 
is given by
\begin{eqnarray}
\label{diff}
\lefteqn{\Delta \hat \sigma = d\hat \sigma^{(0)}\;
\frac{\alpha}{\pi} \frac{1}{2} 
\left\{\log\left(\frac{\hat s}{M_W^2}\right)+1+\frac{\pi^2}{12}
 \right. }
\nonumber\\
& & \left. +(Q_f^2+Q_{f'}^2)\left[
\left(\log\left(\frac{\hat s}{M_W^2}
\delta_s \frac{\delta_{\theta}}{2}\right)+\frac{3}{2} \right)
\; \log\left(\frac{\hat s}{M_W^2}\right) \right. \right.
\nonumber\\
& & \left. \left.+\frac{\pi^2}{6}-2\right]  \right\} \; .
\end{eqnarray}
As we shall see, this difference has a non-negligible effect on the 
transverse mass distribution, and thus on the $W$ boson mass extracted from
experiment.

\section{Numerical Impact on the Transverse Mass Distribution}
\begin{figure}[t]
\epsfbox{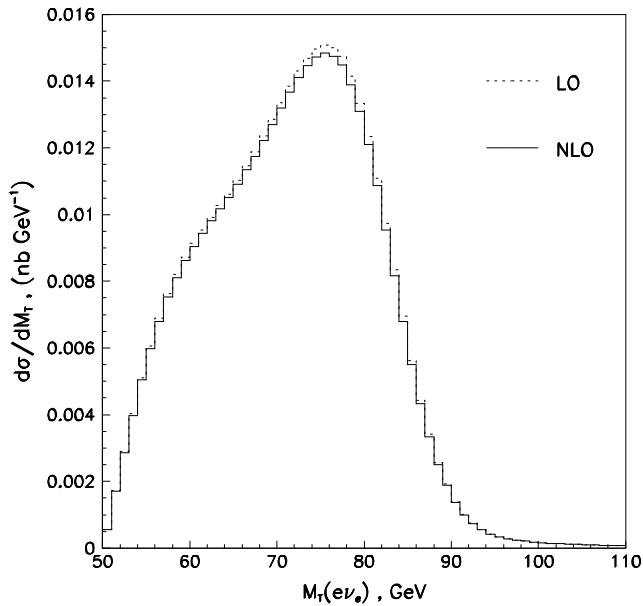}
\caption{The $M_T(e\nu_e)$ distribution of $W$ boson 
production at the Tevatron ($\sqrt{S}=1.8$ TeV).}
\label{fig:mt}
\end{figure}

Since detectors at the Tevatron
cannot directly detect the neutrino produced in the leptonic $W$ boson decay,
$W\to e\nu_e$, and cannot measure the
longitudinal component of the recoil momentum, there is insufficient
information to reconstruct the invariant mass of the $W$ boson. 
Instead, the transverse mass distribution of the final state
lepton pair is used to extract $M_W$.  
In the following, we therefore focus on the $M_T(e\nu_e)$ distribution.
The following acceptance cuts and electron identification 
criteria~\cite{D0Wmass} are used to simulate detector response:
\begin{itemize}
\item
the uncertainty in the energy measurement is simulated by 
Gaussian smearing of the lepton momenta corresponding to the D\O\ 
electromagnetic energy
and missing transverse energy resolution,
\item
the photon and electron are treated as separate particles only if
$\Delta R_{e\gamma}>0.3$. For smaller values of $\Delta R_{e\gamma}$,
the four momentum vectors of the two particles are combined to an
effective electron momentum four-vector (recombination cut). In the region 
$0.2<\Delta R_{e\gamma}<0.4$ the event is rejected if
$E_{\gamma}>0.15 E_e$ (isolation cut),
\item
we impose a cut on the electron transverse energy of 25 GeV, a
missing transverse energy cut of 25 GeV, 
and require the electron pseudorapidity to be $|\eta_e| < 1.2$.
\end{itemize}
\noindent
We use the MRSA set of parton distributions~\cite{mrsa} with 
the factorization and renormalization scales set equal to $M_W$.
For the numerical evaluation of $d \sigma_{2\rightarrow 2}^{initial}$
(see Eq.~(\ref{twoin})) we use the $\overline{\mbox{MS}}$ scheme. 
The leading order
(LO) and next-to-leading order (NLO) $M_T(e\nu)$ differential cross
sections are shown in Fig.~\ref{fig:mt}.  As can be seen, the overall 
effect of the ${\cal O}(\alpha)$ corrections is to reduce the cross section.
\noindent

\begin{figure}[h]
\epsfbox{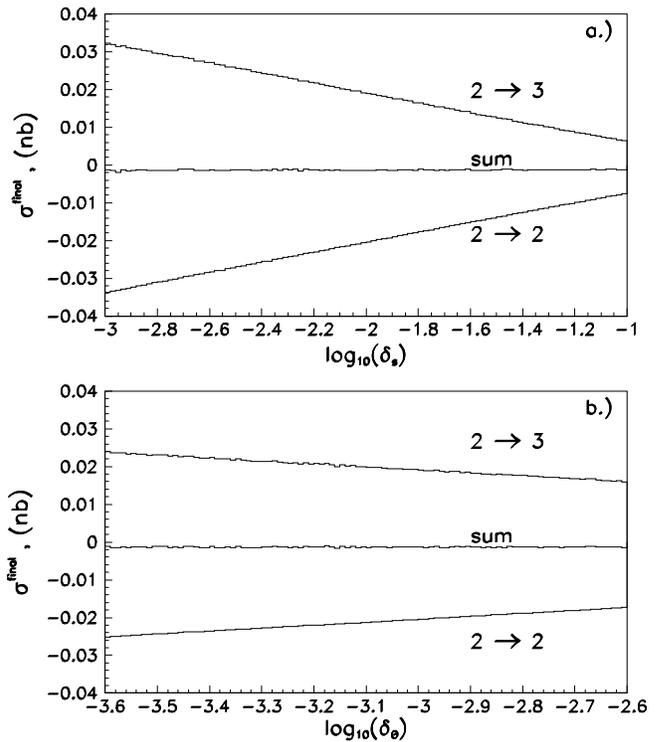}
\caption{The $2\rightarrow 2$ and reduced $2\rightarrow 3$ final state 
contribution a) as a function of the cutoff $\delta_s$ 
with $\delta_{\theta}=0.001$, and b) as a function of 
$\delta_\theta$ with $\delta_s=0.01$.}
\label{fig:ds}
\end{figure}

The dependence on the cutoff parameters $\delta_s$ and $\delta_\theta$ 
must cancel in the sum of the $2\rightarrow 2$ and reduced $2\rightarrow 3$
contributions, provided that the cutoff parameters are chosen sufficiently 
small such that the soft photon and leading-pole approximation
are valid. This is demonstrated in Fig.~\ref{fig:ds} for the final state 
contributions. We have also verified the cancellation for 
the initial state and interference contributions. 
\noindent

\begin{figure}[h]
\epsfbox{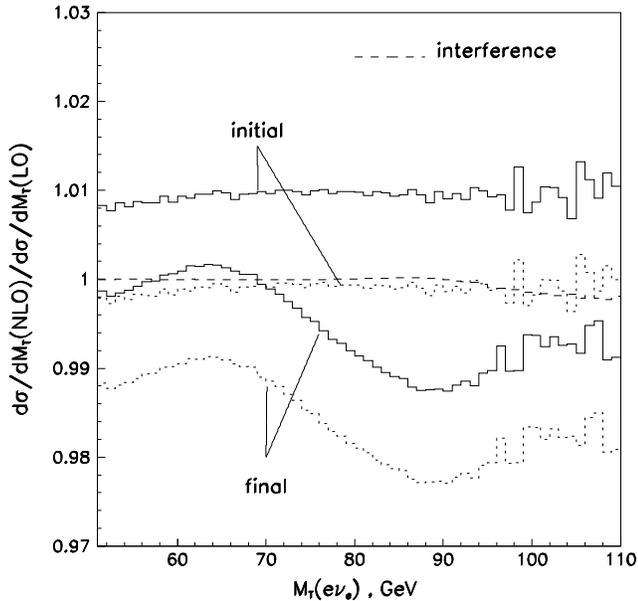}
\caption{The ratio of the NLO to LO $M_T(e\nu_e)$ distribution for
various individual contributions:~the
QED-like initial or final state contributions (solid), the
complete ${\cal O}(\alpha)$ initial and final state 
contributions (short dashed) and the initial--final
state interference contribution (long dashed).}
\label{fig:ratio}
\end{figure}

In Fig.~\ref{fig:ratio}, the effect of the various individual contributions
to the EW ${\cal O}(\alpha)$ corrections on the $M_T(e\nu_e)$
distribution is shown. The following conclusions can be drawn:
\begin{itemize}
\item 
The initial state QED-like contribution uniformly increases  
the cross section by about 1\%, but is largely cancelled 
by the modified weak initial state contribution.  
\item
Both the complete ${\cal O}(\alpha)$ initial state
contribution and the interference contribution
are small and change the shape of the $M_T(e\nu)$ distribution very 
little.  These 
contributions therefore, are expected to have a negligible effect 
on the value of $M_W$ extracted from the data.
\item
The final state QED-like contribution significantly changes 
the shape of the transverse mass distribution and will, therefore, have 
a non-negligible effect on the value of $M_W$ extracted from data.  
As for the initial state, the modified weak final state contribution
reduces the cross section by about $1\%$, but has only a small
effect on the shape of the transverse mass distribution.
\end{itemize}
\noindent
The change in the shape of the $M_T(e\nu_e)$ distribution due to the QED-like 
final state corrections can be easily understood. Photon radiation
reduces the energy of the final state electron and thus the transverse
mass when the electron and photon momenta are not recombined. 

\begin{figure}[h]
\epsfbox{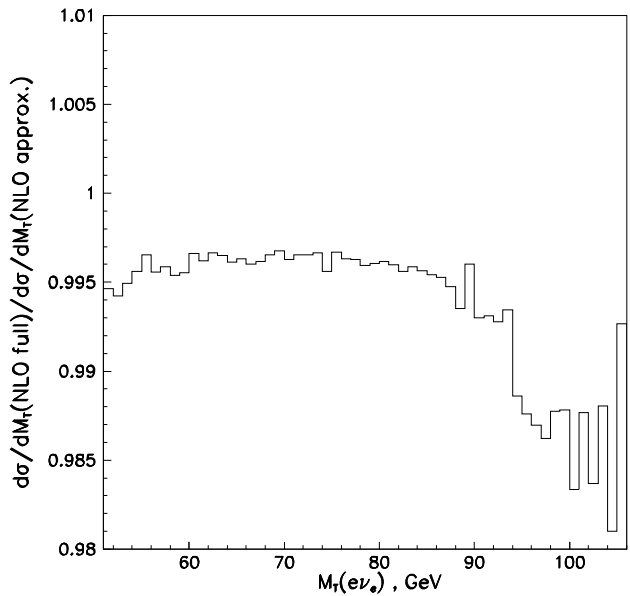}
\caption{The ratio of the complete to the 
approximate NLO $M_T(e\nu_e)$ distribution.}
\label{fig:diff}
\end{figure}

In Fig.~\ref{fig:diff}, we display the ratio of the
$M_T(e\nu_e)$ distribution obtained with our complete NLO calculation 
to the one obtained by using the approximation of Ref.~\cite{berends}.
The dependence of the ratio on the transverse mass is described by 
Eq.~(\ref{diff}). For $M_T(e\nu)<M_W$, most events originate from the
region $\hat s\approx M_W^2$, due to the Breit-Wigner resonance. 
Consequently, there is very little dependence on $M_T$ in that region.
For $M_T(e\nu)>M_W$, the steeply falling cross section in the 
tail of the Breit-Wigner resonance favors events with $\hat s\approx
M_T^2$. Therefore, in this region,
the terms proportional to $\log(\hat s/M_W^2)$ 
in Eq.~(\ref{diff}) cause a change in the shape of the
transverse mass distribution. The difference in the line shape occurs in a 
region of the $M_T(e\nu_e)$ distribution which is sensitive to $M_W$, and 
we expect that the $W$ boson mass extracted when using
the complete calculation will
differ by several MeV from the value obtained using the approximate
calculation.
Since this difference is much smaller than the present
uncertainty for $M_W$, the approximation of Ref.~\cite{berends}
provides an adequate description of $W$ boson production
at the Tevatron for the currently available data.
However, for future precision experiments,
a difference of several MeV in the extracted value of $M_W$ 
can no longer be ignored, and the complete
calculation should be used.

\section{Conclusions}
We have presented the results of a calculation of
the EW ${\cal O}(\alpha)$ corrections to $W$ boson
production at hadron colliders. Both initial and 
(complete) final state corrections, as well as the interference 
between the initial and final state corrections are included in our
calculation. The initial state corrections and the interference 
contribution are found to be small and uniform in $M_T(e\nu_e)$, 
and are expected to have a small effect on the $W$ boson mass extracted from
experimental data.  
As expected, the final state corrections dominate the EW radiative corrections.
They significantly change
the shape of the transverse mass distribution, and thus the value of $M_W$ 
extracted from the data.

We also compared the result of our complete calculation 
with that of Ref.~\cite{berends} which uses an approximation to estimate
the sum of soft and virtual final state corrections. For $M_T>M_W$,
the complete and approximate NLO $M_T(e\nu_e)$ distributions differ
in shape, and we expect that 
they will yield values for the $W$ boson mass which differ by several MeV.

Our calculation substantially improves the treatment of EW radiative 
corrections to $W$ boson production in hadronic collisions,
and will allow to significantly
reduce the systematic uncertainties associated with these
corrections when the $W$ boson mass is extracted from Tevatron data.

\section{Acknowledgements}

We would like to thank M.~Demarteau, S.~Errede and Y-K.~Kim 
for stimulating discussions. This work has been supported in 
part by Department of Energy 
contract No.~DE-AC02-76CHO3000 and NSF grant PHY-9600770.

%

\end{document}